\documentclass[english,twocolumn,english,aip,reprint]{revtex4-1}
\usepackage[T1]{fontenc}
\usepackage[latin9]{inputenc}
\usepackage{color}
\usepackage{babel}
\usepackage{verbatim}
\usepackage{textcomp}
\usepackage{amsmath}
\usepackage{graphicx}
\usepackage[unicode=true,pdfusetitle,
 bookmarks=true,bookmarksnumbered=false,bookmarksopen=false,
 breaklinks=false,pdfborder={0 0 1},backref=false,colorlinks=true]
 {hyperref}
\hypersetup{
 urlcolor=blue,linkcolor=red,citecolor=blue}

\makeatletter
\usepackage{setspace}

\makeatother

\begin{document}
\title{SnO/$\beta$-Ga$_{2}$O$_{3}$ vertical $p$$n$ heterojunction diodes}
\author{Melanie Budde}
\affiliation{Paul-Drude-Institut f\"ur Festk\"orperelektronik, Leibniz-Institut im
Forschungsverbund Berlin e.V., Hausvogteiplatz 5-7, 10117 Berlin,
Germany}
\affiliation{Both authors contributed equally to this work.}
\author{Daniel Splith}
\affiliation{Universit\"at Leipzig, Felix-Bloch-Institut f\"ur Festk\"orperphysik, Halbleiterphysik,
Linn\'{e}stra\ss{}e 5, 04103 Leipzig, Germany}
\affiliation{Both authors contributed equally to this work.}
\author{Piero Mazzolini}
\affiliation{Paul-Drude-Institut f\"ur Festk\"orperelektronik, Leibniz-Institut im
Forschungsverbund Berlin e.V., Hausvogteiplatz 5-7, 10117 Berlin,
Germany}
\affiliation{Present address: Department of Mathematical, Physical and Computer
Sciences, University of Parma, Viale delle Scienze 7/A, 43124 Parma,
Italy}
\author{Abbes Tahraoui}
\affiliation{Paul-Drude-Institut f\"ur Festk\"orperelektronik, Leibniz-Institut im
Forschungsverbund Berlin e.V., Hausvogteiplatz 5-7, 10117 Berlin,
Germany}
\author{Johannes Feldl}
\affiliation{Paul-Drude-Institut f\"ur Festk\"orperelektronik, Leibniz-Institut im
Forschungsverbund Berlin e.V., Hausvogteiplatz 5-7, 10117 Berlin,
Germany}
\author{Manfred Ramsteiner}
\affiliation{Paul-Drude-Institut f\"ur Festk\"orperelektronik, Leibniz-Institut im
Forschungsverbund Berlin e.V., Hausvogteiplatz 5-7, 10117 Berlin,
Germany}
\author{Holger von Wenckstern}
\affiliation{Universit\"at Leipzig, Felix-Bloch-Institut f\"ur Festk\"orperphysik, Halbleiterphysik,
Linn\'{e}stra\ss{}e 5, 04103 Leipzig, Germany}
\author{Marius Grundmann}
\affiliation{Universit\"at Leipzig, Felix-Bloch-Institut f\"ur Festk\"orperphysik, Halbleiterphysik,
Linn\'{e}stra\ss{}e 5, 04103 Leipzig, Germany}
\author{and Oliver Bierwagen}
\affiliation{Paul-Drude-Institut f\"ur Festk\"orperelektronik, Leibniz-Institut im
Forschungsverbund Berlin e.V., Hausvogteiplatz 5-7, 10117 Berlin,
Germany}
\email{bierwagen@pdi-berlin.de}

\date{\today}
\begin{abstract}
\noindent Vertical $pn$ heterojunction diodes were prepared by plasma-assisted
molecular beam epitaxy of unintentionally-doped $p$-type SnO layers
with hole concentrations ranging from $p=10^{18}$ to $10^{19}$\,cm$^{-3}$
on unintentionally-doped $n$-type $\beta$-Ga$_{2}$O$_{3}$(-201)
substrates with an electron concentration of $n=2.0\times10^{17}$\,cm$^{-3}$.
The SnO layers consist of (001)-oriented grains without in-plane expitaxial
relation to the substrate. After subsequent contact processing and
mesa etching (which drastically reduced the reverse current spreading
in the SnO layer and associated high leakage) electrical characterization
by current-voltage and capacitance-voltage measurement was performed.
The results reveal a type-I band alignment and junction transport
by thermionic emission in forward bias. A rectification of $2\times10^{8}$
at $\pm1$\,V, an ideality factor of 1.16, differential specific
on-resistance of 3.9\,m$\Omega\thinspace$cm$^{2}$, and built-in
voltage of 0.96\,V were determined. The $pn$-junction isolation
prevented parallel conduction in the highly-conductive Ga$_{2}$O$_{3}$
substrate (sheet resistance $R_{S}\approx3\thinspace\Omega$) during
van-der-Pauw Hall measurements of the SnO layer on top ($R_{S}\approx150$\,k$\Omega$,
$p\approx2.5\times10^{18}$\,cm$^{-3}$, Hall mobility $\approx1$\,cm$^{2}$/Vs).
The measured maximum reverse breakdown voltage of the diodes was 66\,V,
corresponding to a peak breakdown field 2.2\,MV/cm in the Ga$_{2}$O$_{3}$-depletion
region. Higher breakdown voltages that are required in high-voltage
devices could be achieved by reducing the donor concentration in the
$\beta$-Ga$_{2}$O$_{3}$ to increase the depletion width as well
as improving the contact geometry to reduce field crowding.
\end{abstract}
\maketitle
During the last decade transparent semiconducting oxides (TSOs) have
become a widely investigated class of materials. Their transparency
and wide band gaps are especially suitable for optoelectronic and
power electronic applications. Most studied TSOs are \textit{n}-type
such as Ga$_{2}$O$_{3}$, In$_{2}$O$_{3}$ or SnO$_{2}$. Out of
these, Ga$_{2}$O$_{3}$ with the thermodynamically stable monoclinic
polymorph $\beta$\nobreakdash-Ga$_{2}$O$_{3}$, is predicted to
outperform GaN and SiC for high-voltage power electronics.\citep{Higashiwaki2012,Pearton2018,Galazka2018a}
This advantage is related to its ultra-wide band gap of $E_{g}\simeq$4.8~eV
providing a high break-down field of $\simeq$8~MV/cm and a reasonably
high electron mobility around 200~cm$^{2}$/Vs giving rise to a sufficiently
low on-resistance. In addition, the availability of large area (e.g.,
2 and 4 inch wafers \citep{Galazka2014,Kuramata2016}) bulk $\beta$\nobreakdash-Ga$_{2}$O$_{3}$
single crystals provides the basis for low-defect $\beta$\nobreakdash-Ga$_{2}$O$_{3}$
devices required for ultimate performance.\citep{Ueda1997,Galazka2010,Kuramata2016,Galazka2016}
Edge-termination to manage field crowding in high-voltage power electronic
devices is ideally realized by \textit{pn}-junctions.\citep{Gong2020}
Since bipolar doping is not possible for most TSOs,\citep{Lany_DopingRule,Robertson_DopingRule,Lany2009}
including $\beta$\nobreakdash-Ga$_{2}$O$_{3}$,\citep{Varley2012,Kyrtsos2018}
\textit{n}-type TSOs need to be combined with suitable \textit{p}-type
TSOs to form \textit{pn}-heterojunctions.\citep{Grundmann2016}

The first Ga$_{2}$O$_{3}$-based, all-oxide \textit{pn}-junction
was reported in 2016 by Kokubun et al., combining a $\beta$\nobreakdash-Ga$_{2}$O$_{3}$
single crystal with Li-doped NiO as a \textit{p}-type material.\citep{NiO+Ga2O3}
In the following years \textit{pn}-heterojunction followed with the
\textit{p}-type materials NiO,\citep{PintorMonroy2018,Li2019,Lu2020_Diode,Gong2020}
Cu$_{2}$O,\citep{Watahiki2017} Ir$_{2}$O$_{3}$\citep{Kan2018}
and ZnCo$_{2}$O$_{4}$.\citep{Schlupp2019_Diode} A comparison of
the properties of some of these \textit{pn}-heterojunction diodes
can be found in Ref.\,\citenum{Splith2020}. In the latest publication,
Gong et al. achieved a rectification ratio of the current $I$ at
the voltage $\pm V$ (S$_{V}$$\mathrm{=\frac{I(V)}{I(-V)}}$) of
$S_{3\text{V}}>10^{10}$ and a breakdown voltage (V$_{b}$) of 1.86~kV,
correlated with a maximum ($E_{m}$) electric breakdown field of about
3.5~MV/cm, for NiO/$\beta$\nobreakdash-Ga$_{2}$O$_{3}$ diodes
having a type-II band alignment.\citep{Gong2020} For these devices
an ideality factor ($\eta$) around 2 has been observed,\citep{Gong2020}
whereas values close to 1 have been published, for example by Lu et
al., but with only $S_{V}\approx10^{4}$ for the same materials combination.\citep{Lu2020_Diode}

$\beta$\nobreakdash-Ga$_{2}$O$_{3}$(100)-based diodes with $\eta=1.22$
and $S_{V}>10^{10}$ have been reported by Du et al., using sputtered
SnO$_{x}$ Schottky electrodes. The layers are described as a combination
of SnO and Sn (based on Raman and X-ray photoelectron spectroscopy
measurements) whose resistance between two Ti contacts were determined.
In fact, SnO is a \textit{p}-type oxide with a two orders of magnitude
higher hole mobility than NiO\citep{Zhang2018,Budde2020} and $E_{g}\approx0.7$\,eV.\citep{Ogo2008}
Addressing the metastability of SnO with respect to Sn and SnO$_{2}$
we have determined the growth window for the formation of stoichiometric
SnO by plasma-assisted molecular beam epitaxy (MBE) in our previous
study: SnO was grown by a controlled Sn/O-plasma flux ratio at temperatures
$\leq$400$^\circ$C , resulting in \textit{p}-type conductivity with Hall
hole concentrations ($p_{H}$) and Hall mobilites ranging from $10^{18}$
to $10^{19}$~cm$^{-3}$ and 1 to 6.0 cm$^{2}$/Vs, respectively.
The phase was stable under rapid thermal annealing (RTA) in different
atmospheres up 300$^\circ$C and transformed into $n$-type Sn$_{3}$O$_{4}$
and SnO$_{2}$ at 400$^\circ$C.\citep{Budde2020} 

In this letter, we report the fabrication and characteristics of vertical
\textit{pn}-heterojunction diodes consisting of MBE-grown, $p$-type
SnO layers on unintentionally $n$-type doped $\beta$\nobreakdash-Ga$_{2}$O$_{3}$($\mathrm{\bar{2}01}$)
substrates grown by the edge-defined film-fed growth method (Tamura
corporation).\citep{Kuramata2016} 5~mm$\times$5~mm-large pieces
diced from the $\beta$\nobreakdash-Ga$_{2}$O$_{3}$($\mathrm{\bar{2}01}$)
substrate wafer were etched in phosphoric acid at 130$^\circ$C (removing
$\simeq$500~nm)\citep{Oshima_2009_etching} followed by an annealing
step in oxygen (1~bar) at 950$^\circ$C for 1~hour in a tube furnace to
remove a potentially present polishing-damage layer at the surface
and to regain a stoichiometric Ga$_{2}$O$_{3}$ surface.\citep{Mazzo_etching,Mazzolini2020}
A room-temperature Hall-measurement as described in Ref.\,\citenum{Golz2019}
on a reference piece prepared the same way and from the same Ga$_{2}$O$_{3}$
wafer indicates an electron concentration ($n$) of $2.0\times10^{17}$\,cm$^{-3}$,
taking into account a Hall factor of 1.6.\citep{Ma2016} An ohmic
contact was formed on the substrate backside by electron-beam evaporation
of 20\,nm Ti/100\,nm\,Au and RTA for 1\,minute at 470$^\circ$C in N$_{2}$
atmosphere\citep{Ti_ohmic_anneal,Yao_ohmic_ti,Becky_ohmic} \textit{before
growth} of the SnO layer to prevent its transformation into $n$-type
SnO$_{x}$ during RTA. After that the substrates were in-situ cleaned
using an oxygen plasma {[}0.5~standard cubic centimeters per minute
(sccm), 300\,W{]} at a substrate heater temperature of 400$^\circ$C in the
MBE growth chamber. Two samples (G015 and G016) were grown at 400$^\circ$C
using oxygen fluxes of 0.15~sccm and 0.16~sccm, respectively. For
both runs a plasma power of 300\,W and a metallic Sn flux with a
beam equivalent pressure of 1$\cdot$10$^{-7}$~mbar and a growth
time of 40 minutes were used. A piece of (insulating) c-plane Al$_{2}$O$_{3}$
was co-loaded in each run as a reference samples (A015, A016).

\begin{figure}
\noindent \begin{centering}
\includegraphics[width=8cm]{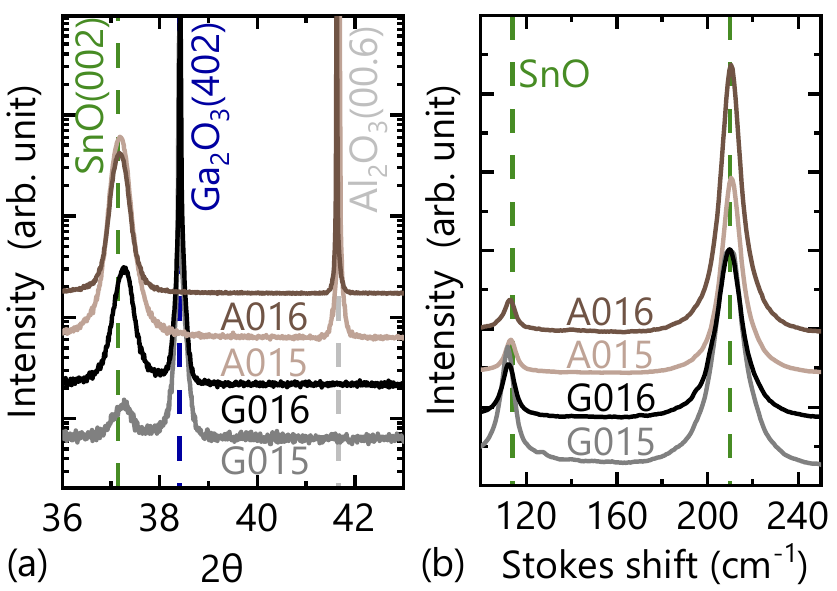}
\par\end{centering}
\caption{(a) XRD 2$\theta$-$\omega$ scans and (b) bulk-sensitive Raman spectra
of the reference samples A015 and A016, as well as the diodes G015
and G016. The Ga$_{2}$O$_{3}$, Al$_{2}$O$_{3}$, and SnO reflexes
are indicated with blue, grey and green dashed lines, respectively,
in (a), and the frequencies of optical phonon lines expected for SnO
are indicated by dashed lines in (b).\citep{Budde2020}\label{fig:XRD}}
\end{figure}
 The results of symmetric, out-of-plane $2\Theta-\omega$ X-ray diffraction
(XRD) using Cu-K$\alpha$ radiation and bulk-sensitive Raman spectroscopy
measurements using an excitation wavelength of 473\,nm, as described
in Ref.\,\citenum{Budde2020}, are shown in Fig.~\ref{fig:XRD}
and confirm the presence of (001)-oriented SnO in all samples. In
contrast to A015 and A016 with defined in-plane epitaxial relation\citep{Budde2020},
the film of G016 shows a random rotational mosaicity as shown in the
supplementary material.\citep{supplement} In-situ thickness measurements
using laser reflectometry\citep{Vogt_2015_SnO2} on the reference
samples A015 and A016 indicate a total SnO thickness of 200\,nm and
170\,nm, respectively. Van-der-Pauw Hall measurements of the SnO
layer of G016 using mesa-isolated (as described below) Greek-cross
structures with Ti/Au top contacts (as described below) revealed $R_{S}\approx150$\,k$\Omega$,
$p\approx2.5\times10^{18}$\,cm$^{-3}$ (assuming a Hall scattering
factor of 1.8),\citep{Hu2019}, and an in-plane Hall mobility $\mu_{H}\approx1$\,cm$^{2}$/Vs.
This result indicates a remarkable $pn$-junction isolation that prevented
parallel conduction in the underlying highly-conductive Ga$_{2}$O$_{3}$
substrate (sheet resistance $R_{S}\approx3\thinspace\Omega$) during
the measurement of the SnO layer on top. In addition, we conducted
Hall measurements in the van-der-Pauw geometry on the reference layers
A015 and A016 for which we expected similar $p$ to those of G015
and G016: A sheet resistance of 184~k$\Omega$ and 46~k$\Omega$
was extracted with $p$=$2.0\times10{}^{18}$~cm$^{-3}$ and $1.8\times10{}^{19}$~cm$^{-3}$
as well as $\mu_{H}$=1.6 and 0.8~cm$^{2}$/Vs ~cm$^{2}$ for A015
and A016, respectively. All measured hole concentrations are well
below the critical value ($p_{\text{Mott}}\approx9\times10^{19}$\,cm$^{-3}$)\citep{Budde2020}
for the Mott transition; the hole mobilities are below those of single
crystalline films due scattering from rotational-domain or grain boundaries
in A015/A016,\citep{Budde2020} or G016, respectively. In the following
we assume the net donor and acceptor concentration $N_{D}$ and $N_{A}$
in the Ga$_{2}$O$_{3}$ and SnO to be equal to the measured $n$
and $p$, respectively.

\begin{figure}
\noindent \begin{centering}
\includegraphics[width=7cm]{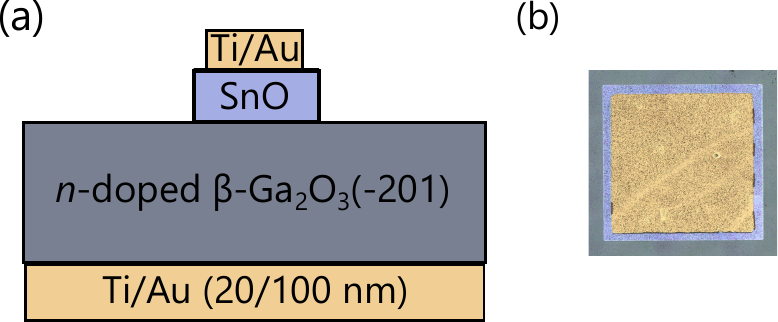}
\par\end{centering}
\vspace{0.3cm}

\noindent \begin{centering}
\includegraphics[width=7cm]{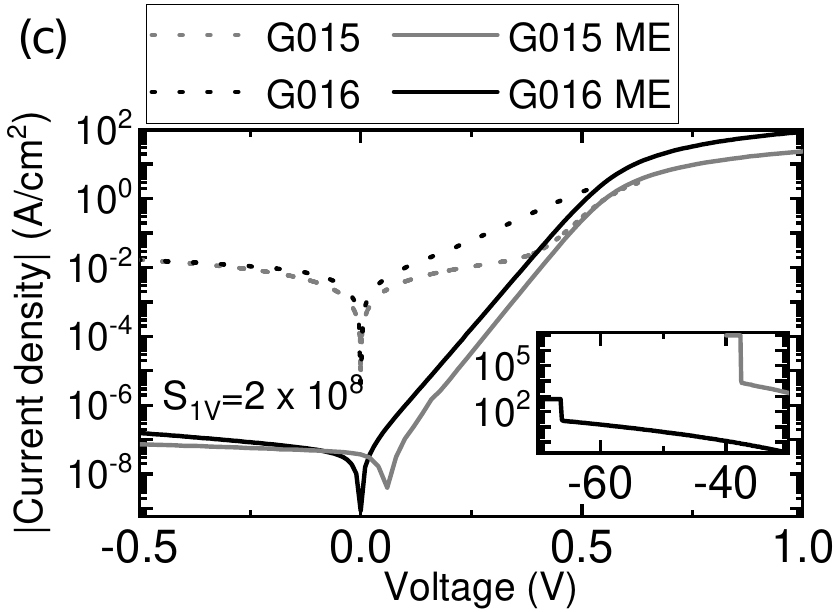}
\par\end{centering}
\caption{(a) Cross-section schematic of the vertical SnO/ Ga$_{2}$O$_{3}$
heterojunction diodes after mesa etching (ME). (b) Micrograph showing
a 180$\times$180~\textmu m$^{2}$ contact pad after ME. The mesa
of the SnO thin film (light blue) is visible around the contact pad
(gold). (c) Room-temperature $IV$ curves of sample G015 and G016
measured on 180$\times$180~\textmu m$^{2}$ contacts in a semi\protect\nobreakdash-logarithmic
plot before and after ME including the rectification factor S$_{V}$
at 1~V. The inset shows the breakdown measurements after ME on one
contact for each sample.\label{fig:IV}}
\end{figure}
\begin{comment}
Die Messung wurde bei G015 an einem 130x130 \textmu m\texttwosuperior{}
kontakt durchgef\"uhrt, ich w\"urde die gr\"osse einfach weglassen. Bei den
normalen Messungen weiss ich grad nicht, welche Messungen f\"ur die Bilder
verwendet wurden.
\end{comment}
{} Square shaped 20~nm Ti/100~nm Au top contacts with sizes varying
between $55\times55$~\textmu m$^{2}$ and $180\times180$~\textmu m$^{2}$
were defined on the grown SnO layer using a photolithography, electron-beam
evaporation, and lift-off without additional RTA to prevent the transformation
of the SnO layer into $n$-type SnO$_{x}$.\citep{Budde2020} Notwithstanding,
these contacts are ohmic with specific contact resistance of $\rho_{c}\approx0.05$\,m$\Omega\thinspace$cm$^{2}$
on A015 and $\rho_{c}\approx3.4$\,m$\Omega\thinspace$cm$^{2}$
on G016 as shown in the supplementary material.\citep{supplement}
After initial current-voltage ($IV$) measurements between top contacts
and bottom contact, mesa etching of the SnO layer was performed on
G015 and G016 to isolate the top contacts using a lithographically-defined
resist mask and an inductively coupled plasma (ICP) inside a reactive
ion etching system. A gas combination of 5~sccm Cl$_{2}$ and 20~sccm
BCl$_{3}$ at a pressure of 1.3~Pa resulted in a suitable etch rate
of about 45~nm/min at an ICP coil power of 100~W and a DC bias power
of 25~W. Fig.~\ref{fig:IV} (a) schematically illustrates the cross-section
of the resulting diode. A top-view micrograph of a 180$\times$180~\textmu m$^{2}$
contact pad after mesa etching (ME) is shown in Fig.~\ref{fig:IV}
(b). Fig.~\ref{fig:IV} (c) shows typical room-temperature (RT) current-voltage
($IV$) characteristics of the two \textit{pn}\nobreakdash-junctions
before and after mesa etching with voltage applied to the $180\times180$~\textmu m$^{2}$
top contacts and the grounded bottom contact. In all cases rectification
as expected for a \textit{pn}-diode can be observed. Before mesa
etching, however, a high reverse current is observed for both samples,
resulting in a $S_{\text{1V}}\approx100$. A drastic reduction of
the reverse current was achieved by mesa etching, resulting in a significant
increase of $S_{\text{1V}}$ to $2\times10{}^{8}$. This improvement
can be explained by preventing the spreading of the reverse current
in the SnO layer over the entire sample area as detailed in the supplementary
information.\citep{supplement} The inset of Fig.~\ref{fig:IV}
(c) shows the breakdown behavior of G015 and G016. Breakdown voltages
of -37~V and -66~V were measured on one contact of G015 and G016,
respectively. The heterojunction $IV$ characteristics in forward
direction can be described by the Shockley equation:\citep{Shockley1949,Schlupp2019_Diode}

\begin{equation}
I=I_{S}\left[\exp\left(\frac{qV-IR_{S}}{\eta k_{B}T}\right)-1\right]+\frac{V-IR_{s}}{R_{p}}+I_{0}.\label{eq:Diode Eq}
\end{equation}

\begin{figure}[h]
\includegraphics[width=7cm]{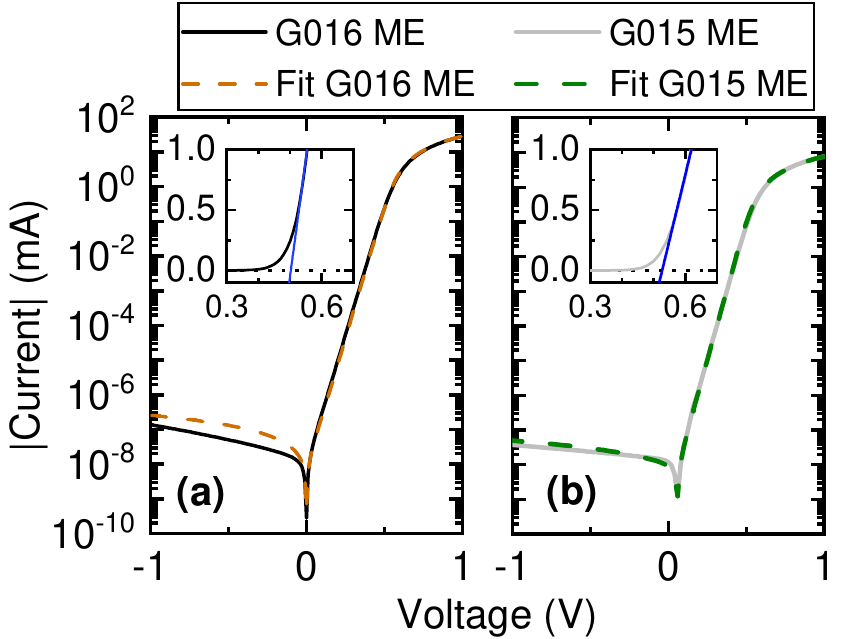}\caption{Room-temperature $IV$ curves measured on $180\times180$~\textmu m$^{2}$
contact pads after ME on the samples (a) G015 (grey) and (b) G016
(black) including the modeled curve (green and brown) using the Shockley
equation. The inset shows the measurement of each sample in the linear
plot together with the fit (blue) of the turn-on voltage.\label{fig:IV-FIT}}
\end{figure}

\noindent Here, $I_{S}$ is the saturation current, $\eta$ the ideality
factor, $k_{B}$ the Boltzmann constant, $q$ the elementary charge,
$T$ the absolute temperature, $V$ the applied voltage, and $R_{s}$
and $R_{p}$ are the series and parallel resistance, respectively.
In contrast to G016 ME, G015 ME shows a slight shift towards positive
voltages. This shift is caused by a capacitive charging current during
the voltage sweep which can be described by $I_{0}$.\citep{Grundmann2018}
The modeled curves together with the measured data are shown in Fig.~\ref{fig:IV-FIT}
for two diodes. In both cases $\eta$ is close to unity, 1.06 and
1.16 for G015 ME and G016 ME, respectively, which indicates a high
diode quality. %
\begin{comment}
\noindent For G016 ME the reverse bias can not be described by the
Shockley equation, indicating a different transport behavior in this
voltage region as a result of the higher hole density, most likely
a tunneling current. Nevertheless, $\eta$ and $R_{s}$ are both extracted
in the forward region.
\end{comment}
{} The turn-on voltage was found to be between 0.50~V and 0.52~V by
a linear fit of the forward bias (see inset of Fig~\ref{fig:IV-FIT}).
From the series resistance determined by the fit, the differential
specific on-resistance was obtained to be 16.5~m$\Omega$\,cm$^{2}$
and 3.9~m$\Omega$\,cm$^{2}$ for G015 ME and G016 ME, respectively.

\noindent 
\begin{figure}
\includegraphics[width=3.2cm]{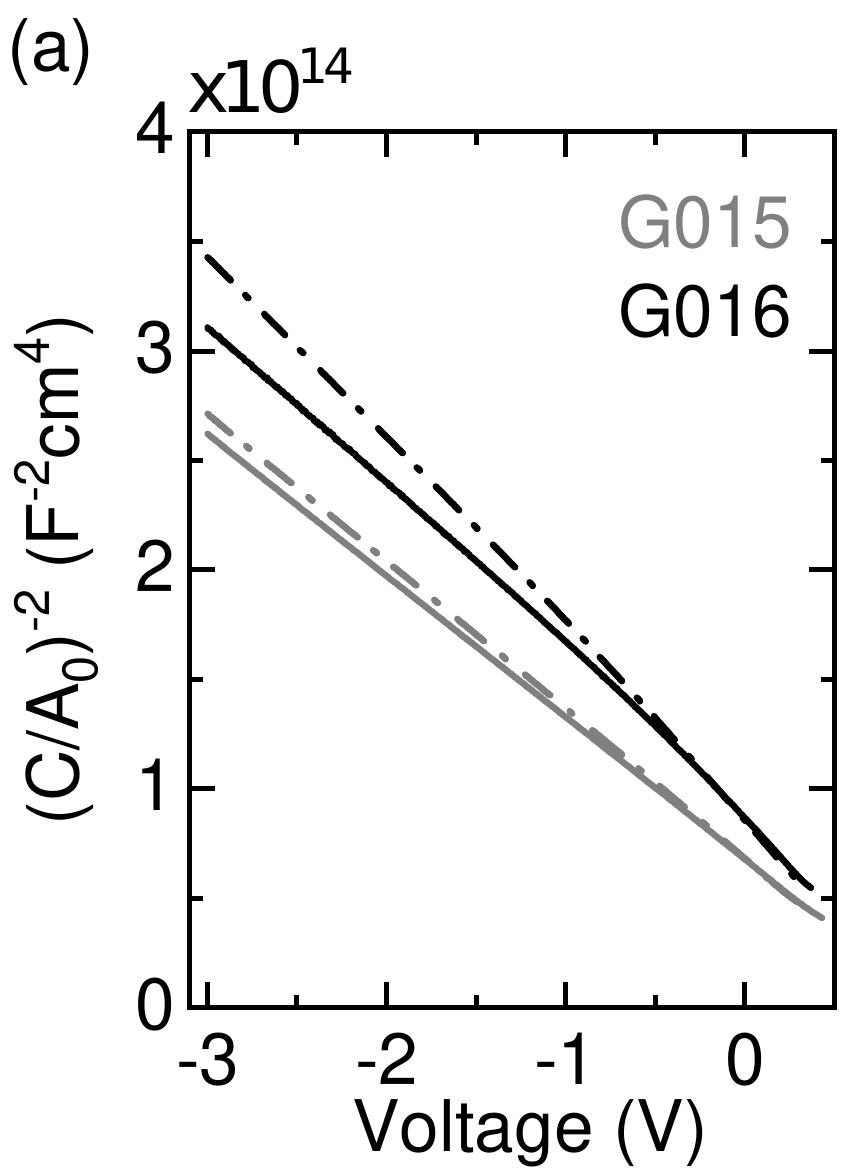}\includegraphics[width=5.3cm]{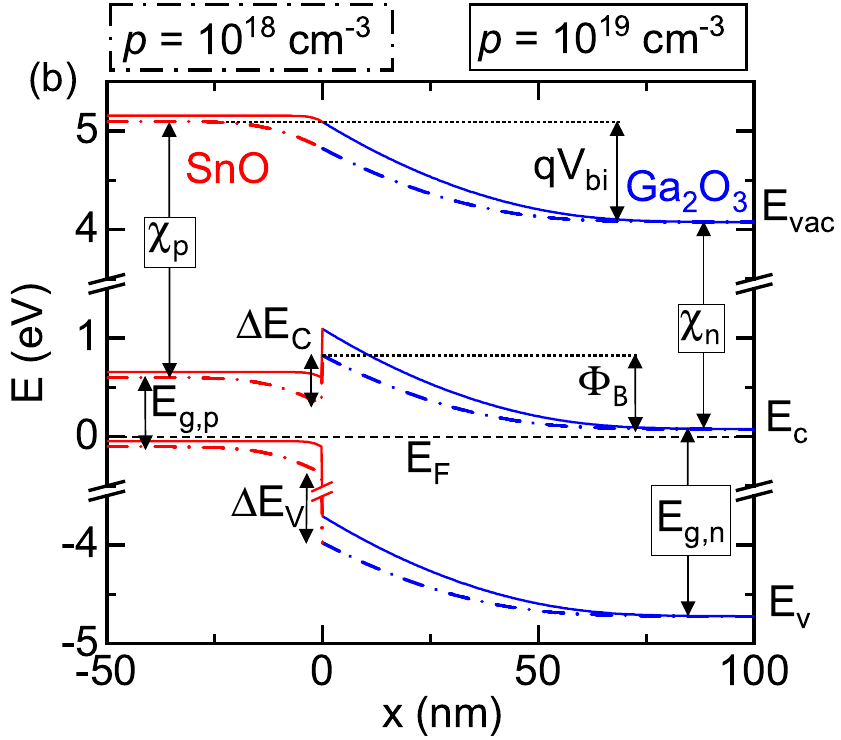}\caption{(a) Results of the $CV$ measurements of two contacts on each, G015
and G016 represented by $(C/A_{0})^{-2}(V)$ plot with contact area
$A_{0}$. (b) Estimated band diagram of the $p$-SnO/$n$-Ga$_{2}$O$_{3}$
junction using values from literature and the determined built-in
potential. The diagram was calculated using the Open Band Parameters
Device Simulator package.\citep{obpds}\label{fig:CV}}
\end{figure}
Room-temperature capacitance-voltage ($CV$) measurements at 1~MHz
are shown in Fig.~\ref{fig:CV}(a) as $C^{-2}-V$ plot for both samples
after ME. The capacitance of a \textit{pn}-heterojunction is described
by\citep{Donnelly1967_CV}
\begin{equation}
C=A_{0}\left[\frac{q\varepsilon_{n}\varepsilon_{p}\varepsilon_{0}N_{D}N_{A}}{2(\varepsilon_{n}N_{D}+\varepsilon_{p}N_{A})}\right]^{1/2}(V_{bi}-V-k_{B}T)^{-1/2}.
\end{equation}

\noindent The relative dielectric constants of SnO, Ga$_{2}$O$_{3}$
and the permittivity of vacuum are $\varepsilon_{p}$, $\varepsilon_{n}$
and $\varepsilon_{0}$, respectively. $V_{bi}$, $A_{0}$, $N_{D}$
and $N_{A}$ are the built-in potential, the area of the junction,
the net concentration of donors in Ga$_{2}$O$_{3}$ and that of acceptors
in SnO, respectively. Dielectric constants $\varepsilon_{p}$ and
$\varepsilon_{n}$ of 18.8 for SnO\citep{Li_perm_SnO} and 10 for
Ga$_{2}$O$_{3}$ \citep{Fiedler2019} were used. The built-in potential
can be extracted by a linear extrapolation of $C^{-2}$ to zero whose
slope yields the effective net doping density $N_{t}=\frac{1}{\varepsilon_{t}}\frac{\varepsilon_{n}N_{D}\varepsilon_{p}N_{A}}{\varepsilon_{n}N_{D}+\varepsilon_{p}N_{A}}$.
Due to the significantly higher $N_{A}$ of the SnO layer than $N_{D}$
of the Ga$_{2}$O$_{3}$ substrate, the depletion region mainly develops
inside the Ga$_{2}$O$_{3}$ whose properties are thus expected to
dominate $N_{T}$: Assuming $\varepsilon_{t}=\varepsilon_{n}$ yields
$N_{t}=(2.1\pm0.1)\times10^{17}$\,cm$^{-3}$, which is in good quantitative
agreement with $n=2.0\times10^{17}$\,cm$^{-3}$ in the Ga$_{2}$O$_{3}$.
No significant differences can be seen for the values of the two samples
so that we report the average value extracted from two measured diodes
on each sample, resulting in $V_{bi}=1.07\pm0.03\,\mathrm{V}.$ Using
$V_{bi}$ and known material parameters of Ga$_{2}$O$_{3}$ {[}band
gap $E_{\mathrm{g,n}}=4.8\,\mathrm{eV}$, electron affinity $\chi_{\mathrm{n}}=4.0\,\mathrm{eV}$,\citep{Mohamed2012}
$\varepsilon_{\mathrm{n}}=10$, density-of-states effective electron
mass $m_{\mathrm{eff,e}}=0.28\,m_{0}$\citep{Knight2018} ($m_{0}$
is the free electron mass){]}, and SnO (band gap $E_{\mathrm{g,p}}=0.7\,\mathrm{eV}$
,\citep{Ogo2008} $\varepsilon_{\mathrm{p}}=18.8$, density-of-states
effective hole mass $m_{\mathrm{eff,h}}=1.7\,m_{0}$\citep{Varley2013}),
we can calculate an estimated band alignment diagram, shown in Fig~\ref{fig:CV}(b)
for $N_{\mathrm{D,n}}=2\times10^{17}\,\mathrm{cm^{-3}}$and $N_{\mathrm{A,p}}=1\times10^{18}\,\mathrm{cm^{-3}}$
as well as $N_{\mathrm{A,p}}=1\times10^{19}\,\mathrm{cm^{-3}}$. The
higher acceptor concentration changes the band diagram slightly (decrease
of the band bending in the $p$-region, increase in the $n$-region)
since the depletion at the heterointerface spreads predominantly to
the substrate. In order to obtain a built-in potential of $1.1\,\mathrm{V}$,
it is necessary to assume an electron affinity of $\chi_{\mathrm{p}}=4.5\,\mathrm{eV}$
for SnO. This value is in the same range as values reported in literature,
which scatter from $3.59\,\mathrm{eV}$ to $5.1\,\mathrm{eV}$.\citep{Xu2000,Li2015,Ogo2009}

\noindent The band alignment shown in Fig.\,\ref{fig:CV}(b) is a
type-I alignment.\citep{Grundmann2016} Due to the small band-gap
of SnO, the conduction band of SnO is below that of Ga$_{2}$O$_{3}$.
This renders thermionic emission as a possible transport mechanism
for the diode; above the conduction band, thermally accessible states
exist, similar to a Schottky barrier diode. Electrons injected from
the Ga$_{2}$O$_{3}$ can either drift through the SnO to the Ohmic
metal contact or recombine in the SnO.%
\begin{comment}
\noindent (which is unlikely due to the indirect nature of the bandgap
of SnO?)
\end{comment}
{} Temperature dependent $IV$-measurements between $50\,\mathrm{K}$
and $380\,\mathrm{K}$ verify thermionic emission as the dominating
transport mechanism in forward direction:
\begin{figure}
\noindent \begin{centering}
\includegraphics[width=7cm]{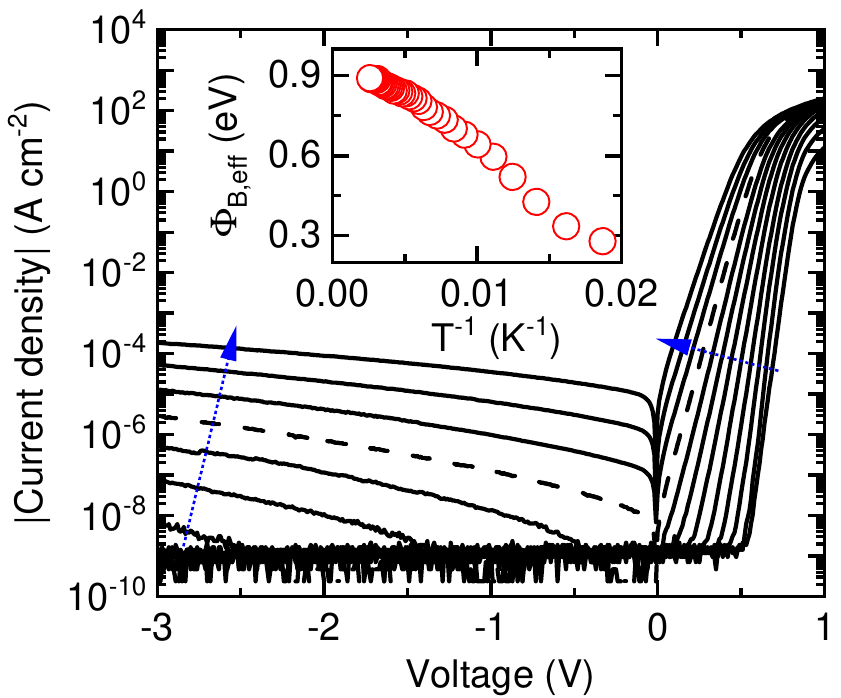}
\par\end{centering}
\caption{Temperature dependent $IV$-characteristics of one SnO/Ga$_{2}$O$_{3}$
contact for selected temperatures. The dashed line is the measurement
at $T=300\,\mathrm{K}$, each characteristic has a temperature difference
of about $30\,\mathrm{K}$ to its neighbors. At $|j|\approx10^{-9}\,\mathrm{A\,cm^{-2}}$,
the noise level of the measurement unit is reached. In the inset,
a plot of the effective barrier height in dependence on $T^{-1}$
is shown. Dotted arrows indicate the trend with increasing $T$.}
\label{fig:IVT}
\end{figure}
 In Fig~\ref{fig:IVT}, an example of the resulting $IV$-characteristics
for selected temperatures is shown. Similar characteristics were observed
also for other contacts. By fitting with the Shockley equation, we
determined $I_{S}$. For thermionic emission over a laterally homogeneous
barrier $\log(I_{\mathrm{S}}/T\text{\texttwosuperior})$ is supposed
to be linear with $T^{-1}$. Instead, we found a quadratic behavior
for $T>100\,\mathrm{K}$ (see Fig.\,S6 in the supplementary material\citep{supplement})
which indicates thermionic emission over a Gaussian distributed lateral
inhomogeneous barrier.\citep{Werner1991} Therefore, the effective
barrier height $\Phi_{\mathrm{B,eff}}$ was calculated from the saturation
current using 
\[
I_{\mathrm{S}}=A_{0}A^{*}\frac{m_{\mathrm{eff,n}}}{m_{\mathrm{0}}}T\text{\texttwosuperior}\exp\left(-\frac{\Phi_{\mathrm{B,eff}}}{k_{\mathrm{B}}T}\right)\;,
\]
where $A^{*}$ is the Richardson constant. At $T=300\,\mathrm{K}$,
a value of $\Phi_{\text{B,eff}}=0.9\,\mathrm{eV}$ was determined.
By plotting $\Phi_{\mathrm{B,eff}}$ vs. $T^{-1}$ (see inset of Fig.~\ref{fig:IVT})
the mean barrier height $\overline{\Phi}{}_{\mathrm{B,0}}=1.02\pm0.04\,\mathrm{eV}$
was determined as mean value of the linear extrapolation of $\Phi_{\mathrm{B,eff}}$
to $T^{-1}=0$ for four different contacts.\citep{Werner1991} The
good agreement between the mean barrier height and the built-in voltage
corroborates the assumption of dominating thermionic emission. More
details on the parameter extraction of the inhomogeneous barrier are
given in the supplementary material.\citep{supplement}

In conclusion, we have demonstrated that SnO (with $p=10^{18}$ to
$10^{19}$\,cm$^{-3}$) and $\beta$-Ga$_{2}$O$_{3}$ (with $n=2\times10^{17}$\,cm$^{-3}$)
form a rectifying $pn$-heterojunction with a type-I band alignment
and junction transport in forward bias by thermionic emission. A rectification
of $2\times10^{8}$ at $\pm1$\,V, ideality factor of 1.16, differential
specific on-resistance of 3.9\,m$\Omega\thinspace$cm$^{2}$, built-in
voltage of 0.96\,V, and reverse breakdown voltage of 66\,V were
achieved. From the band diagram, the depletion layer width $w_{d}=610$\,nm
and maximum breakdown field $E_{m}=2.2$\,MV/cm were estimated at
the breakdown voltage. This $E_{m}$ is appreciable but well below
the theoretical limit of Ga$_{2}$O$_{3}$ (8\,MV/cm),\citep{Higashiwaki2012}
likely related to field spikes at the corners of the square shaped
contacts as well as missing field plates. By reducing the donor concentration
of the Ga$_{2}$O$_{3}$ (to increase the depletion layer thickness)
and improving the contact geometry the breakdown voltage can be increased
towards values required in high-voltage devices. For example, a donor
concentration of $\approx10^{16}$\,cm$^{-3}$ while maintaining
a comparably high $p$ in the SnO to keep the (high-field) depletion
region in the Ga$_{2}$O$_{3}$ could enable breakdown voltages of
$\approx1$\,kV.\citep{Splith2020}
\begin{acknowledgments}
\noindent We would like to thank H.-P. Sch\"onherr, C. Stemmler and
K. Morgenroth for MBE support, S. Rauwerdink, W. Anders, and W. Seidel
for sample processing, and Y. Takagaki for critically reading the
manuscript. This work was performed in the framework of GraFOx, a
Leibniz-ScienceCampus partially funded by the Leibniz association.
M.B. and J.F. gratefully acknowledge financial support by the Leibniz
association. D.S., H.v.W, and M.G. acknowlegde funding from the European
Social Fund within the Young Investigator Group \textquotedblleft Oxide
Heterostructures\textquotedblright{} (SAB 100310460) as well as support
by Universit\"at Leipzig within the research profile area \textquotedblleft Complex
matter\textquotedblright .
\end{acknowledgments}

\subsection*{DATA AVAILABILITY}

The data that support the findings of this study are available from
the corresponding author upon reasonable request.

\bibliographystyle{bibs/unsrtnm_JAPL_new}
\bibliography{bibs/Diodes,bibs/GaN+NiO_paper,bibs/thesis_bib}

\end{document}

\title{Supplementary Material to ---\\
SnO/$\beta$-Ga$_{2}$O$_{3}$ vertical $p-n$ heterojunction diodes}
\author{Melanie Budde}
\affiliation{Paul-Drude-Institut f\"ur Festk\"orperelektronik, Leibniz-Institut im
Forschungsverbund Berlin e.V., Hausvogteiplatz 5-7, 10117 Berlin,
Germany}
\affiliation{Both authors contributed equally to this work.}
\author{Daniel Splith}
\affiliation{Universit\"at Leipzig, Felix-Bloch-Institut f\"ur Festk\"orperphysik, Halbleiterphysik,
Linn\'{e}stra\ss{}e 5, 04103 Leipzig, Germany}
\affiliation{Both authors contributed equally to this work.}
\author{Piero Mazzolini}
\affiliation{Paul-Drude-Institut f\"ur Festk\"orperelektronik, Leibniz-Institut im
Forschungsverbund Berlin e.V., Hausvogteiplatz 5-7, 10117 Berlin,
Germany}
\affiliation{Present address: Department of Mathematical, Physical and Computer
Sciences, University of Parma, Viale delle Scienze 7/A, 43124 Parma,
Italy}
\author{Abbes Tahraoui}
\affiliation{Paul-Drude-Institut f\"ur Festk\"orperelektronik, Leibniz-Institut im
Forschungsverbund Berlin e.V., Hausvogteiplatz 5-7, 10117 Berlin,
Germany}
\author{Johannes Feldl}
\affiliation{Paul-Drude-Institut f\"ur Festk\"orperelektronik, Leibniz-Institut im
Forschungsverbund Berlin e.V., Hausvogteiplatz 5-7, 10117 Berlin,
Germany}
\author{Manfred Ramsteiner}
\affiliation{Paul-Drude-Institut f\"ur Festk\"orperelektronik, Leibniz-Institut im
Forschungsverbund Berlin e.V., Hausvogteiplatz 5-7, 10117 Berlin,
Germany}
\author{Holger von Wenckstern}
\affiliation{Universit\"at Leipzig, Felix-Bloch-Institut f\"ur Festk\"orperphysik, Halbleiterphysik,
Linn\'{e}stra\ss{}e 5, 04103 Leipzig, Germany}
\author{Marius Grundmann}
\affiliation{Universit\"at Leipzig, Felix-Bloch-Institut f\"ur Festk\"orperphysik, Halbleiterphysik,
Linn\'{e}stra\ss{}e 5, 04103 Leipzig, Germany}
\author{and Oliver Bierwagen}
\affiliation{Paul-Drude-Institut f\"ur Festk\"orperelektronik, Leibniz-Institut im
Forschungsverbund Berlin e.V., Hausvogteiplatz 5-7, 10117 Berlin,
Germany}
\email{bierwagen@pdi-berlin.de}
\date{\today}

\maketitle
\renewcommand{\thefigure}{S\arabic{figure}}

\section*{XRD analysis of the SnO film on $\beta$-Ga$_{2}$O$_{3}$(-201)
in sample G016}

\begin{figure*}
\includegraphics[width=15cm]{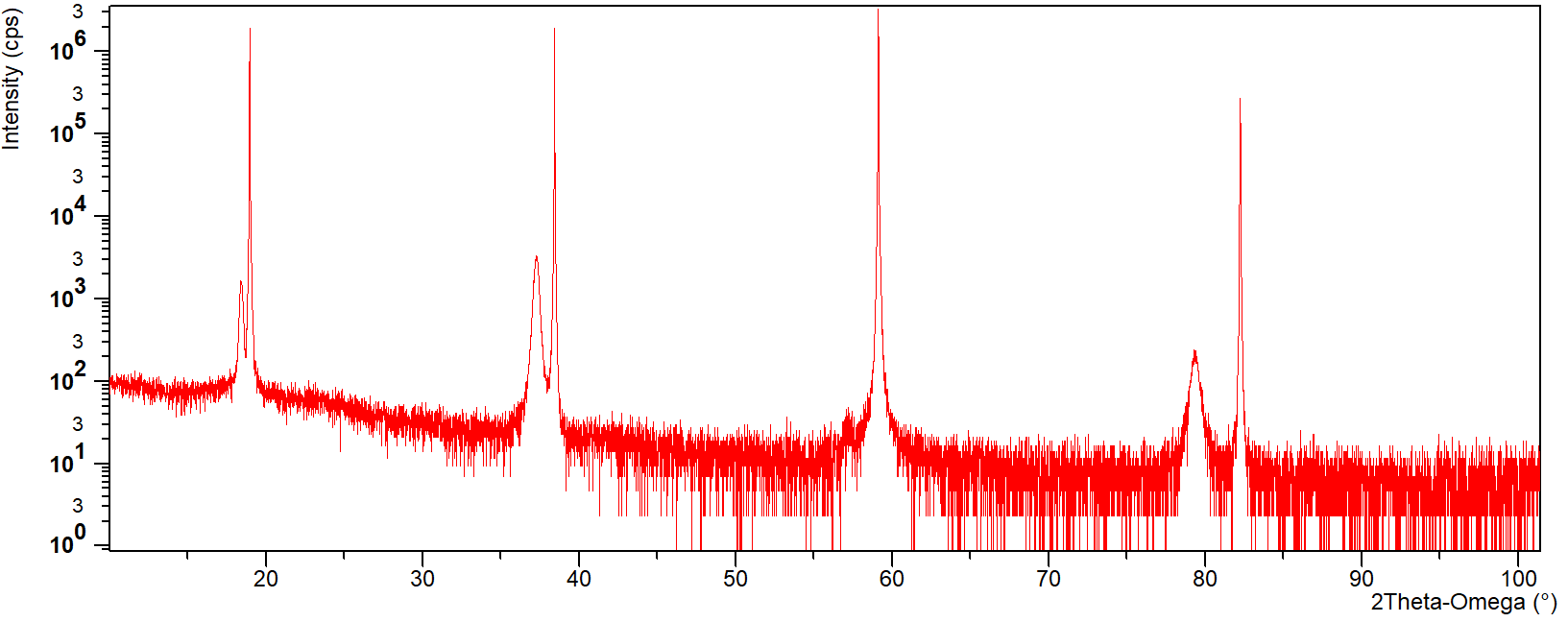}\caption{XRD wide-range out-of-plane symmetric $2\Theta-\omega$ scan of sample
G016.\label{fig:S2thom}}
\end{figure*}
A wide-range out-of-plane symmetric $2\Theta-\omega$ scan of G016
shown in Fig.\,\ref{fig:S2thom} indicates the presence of different
orders of the $\beta$-Ga$_{2}$O$_{3}$-201 reflex (sharp, strong
reflexes) accompanied on the left hand side by different orders of
the SnO 002 reflex (weaker and wider). These results confirm the presence
of out-of-plane 001 oriented SnO. 
\begin{figure*}
\includegraphics[height=6.5cm]{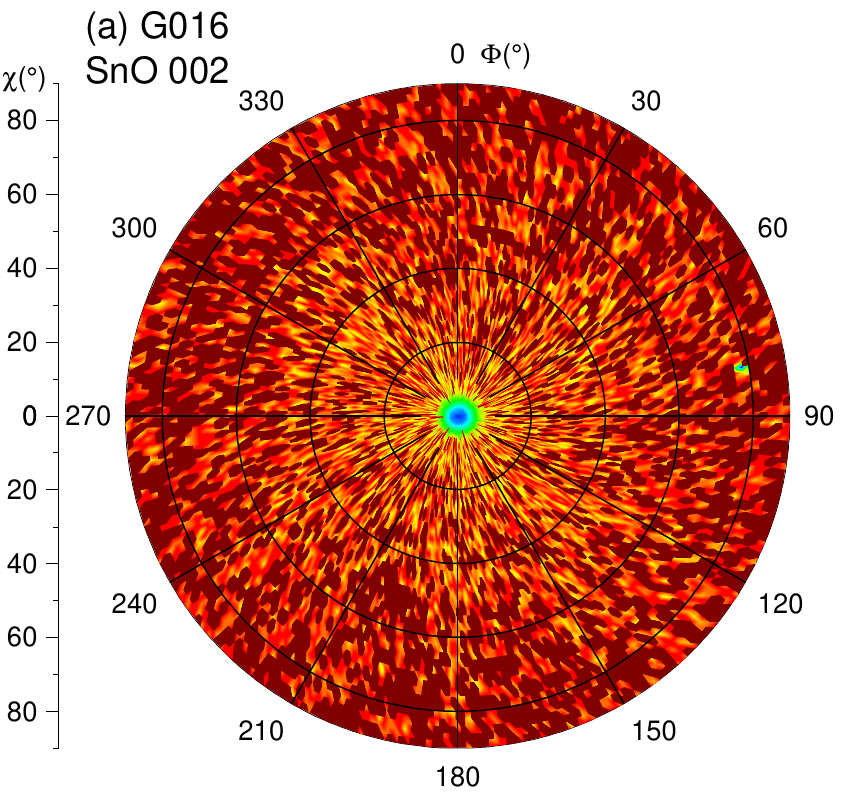}\hspace{0.5cm}\includegraphics[height=6.5cm]{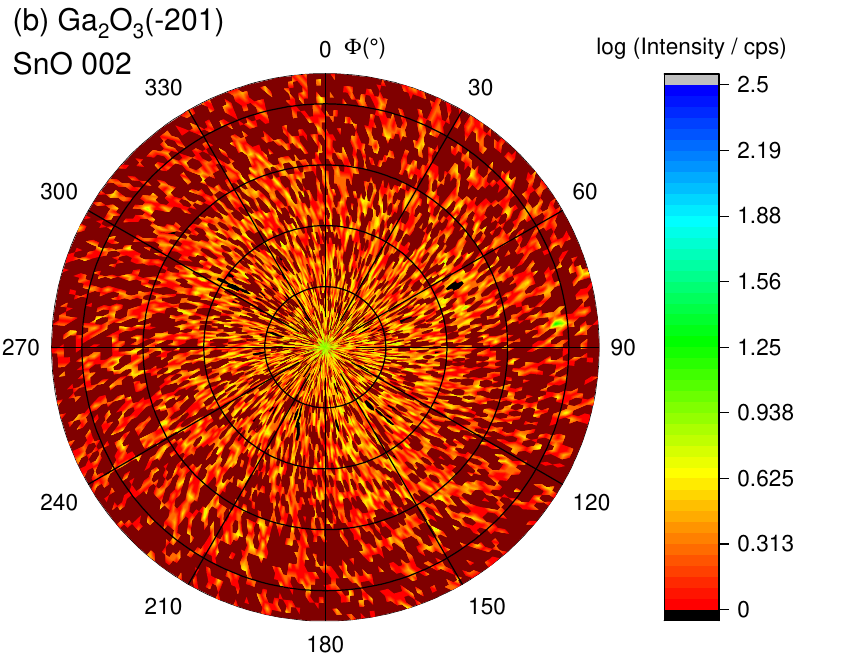}

\vspace{0.5cm}

\includegraphics[height=6.5cm]{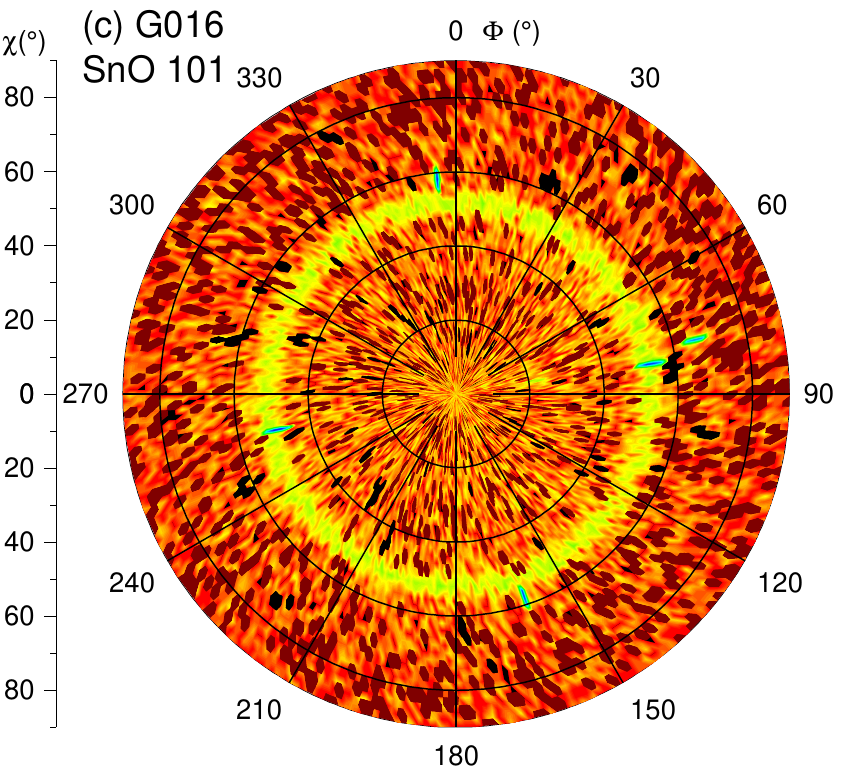}\hspace{0.5cm}\includegraphics[height=6.5cm]{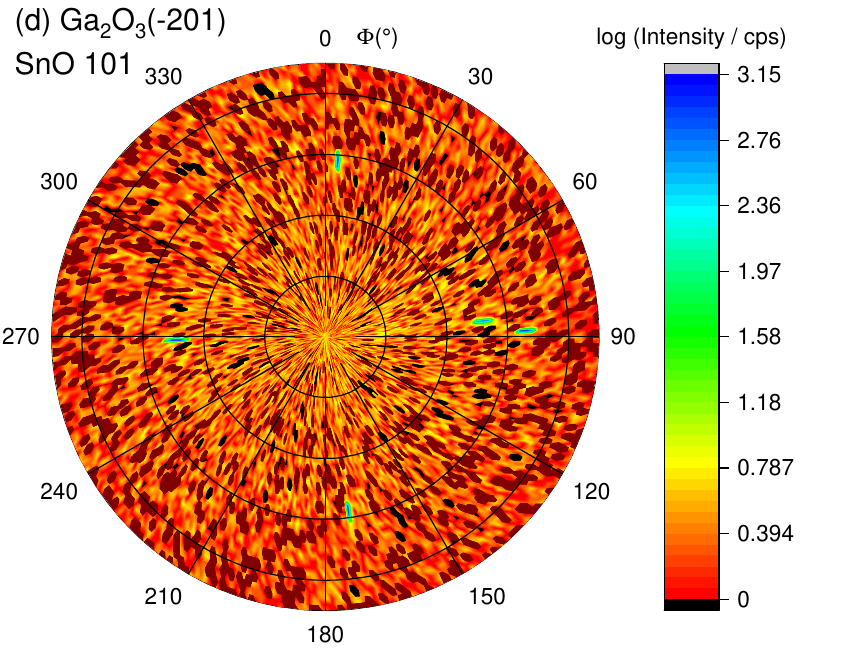}\caption{XRD texture maps of the SnO 002 (a,b) and SnO 101 (c,d) reflex measured
on sample G016 (a, c) and a Ga$_{2}$O$_{3}$(-201) substrate (b,
d) as reference.\label{fig:STexture}}
\end{figure*}
Texture maps, shown in Fig.\,\ref{fig:STexture}, of the 002 and
101 reflex of SnO were measured on G016 as well as a $\beta$-Ga$_{2}$O$_{3}$(-201)
substrate for reference. The map of the SnO 002 reflex shown in Fig.\,\ref{fig:STexture}(a)
confirms the out-of-plane 001 orientation of the SnO film by the strong
intensity of surface parallel SnO(001) planes (at $\chi=0{^\circ}$).
The additonal intensity visible at (at $\chi\approx80{^\circ},\phi\approx80{^\circ}$)
is also present in the texture map on a $\beta$-Ga$_{2}$O$_{3}$(-201)
substrate without grown SnO film (Fig.\,\ref{fig:STexture}(b)) and
is thus not related to the SnO film. 

The in-plane orientation of the SnO film can be seen in the texture
map of the SnO 101 reflex shown in Fig.\,\ref{fig:STexture}(c).
The same type of map for the $\beta$-Ga$_{2}$O$_{3}$(-201) substrate
without grown SnO film is shown in Fig.\,\ref{fig:STexture}(d) for
reference. Again, reflexes related to the substrate are visible in
G016. At $\chi\approx52{^\circ}$, the tilt angle of the SnO 101 plane
with respect to the 001 plane, a weak ring shaped pattern is visible
indicating a polycrystalline SnO layer with randomly rotated but 001
out-of-plane oriented grains. 
\begin{figure}
\includegraphics[width=8cm]{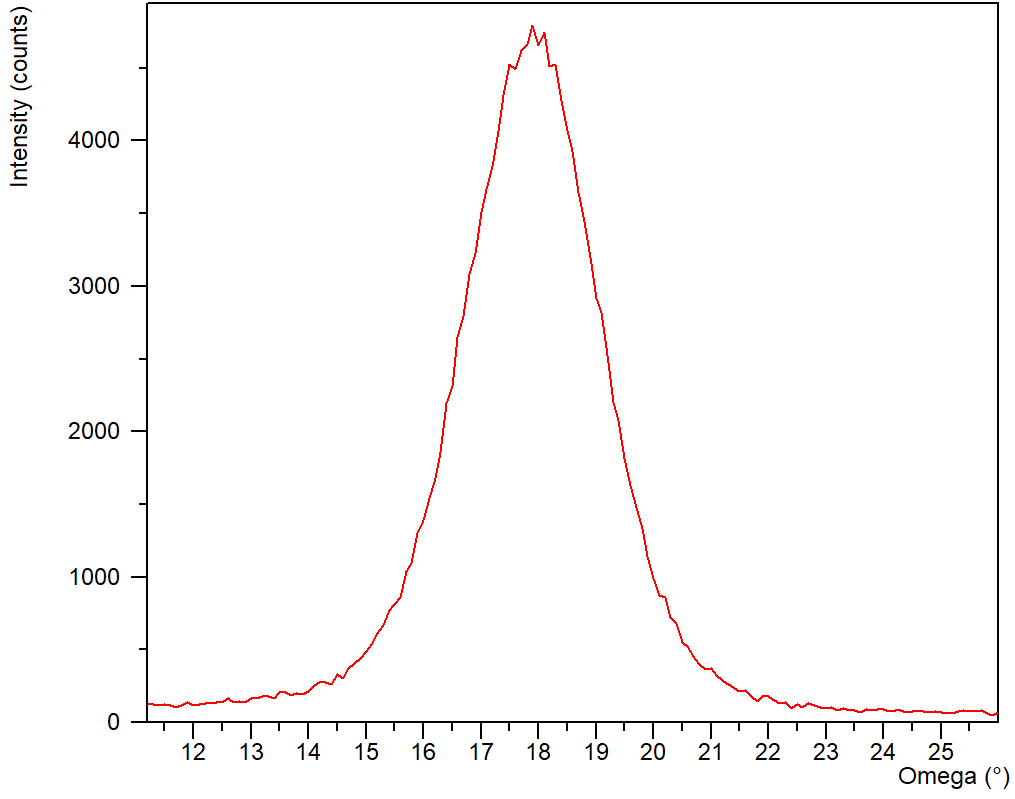}\caption{XRD $\omega$-rocking curve of the SnO 002 reflex in sample G016.\label{fig:Som}}
\end{figure}
The $\omega$-rocking curve of the SnO 002 reflex, shown in Fig.\,\ref{fig:Som},
exhibits a full-width-at-half-maximum of 2.7$^\circ$, indicative of the tilt
mosaic of the out-of-plane orientation.

Thus, the SnO film conists of 001 oriented grains without any in-plane
expitaxial relation to the $\beta$-Ga$_{2}$O$_{3}$(-201) substrate.\clearpage{}

\section*{Ohmic Ti/Au contacts to $p$-type SnO on sampleS A015 and G016}

\begin{figure*}
\includegraphics[width=6.5cm]{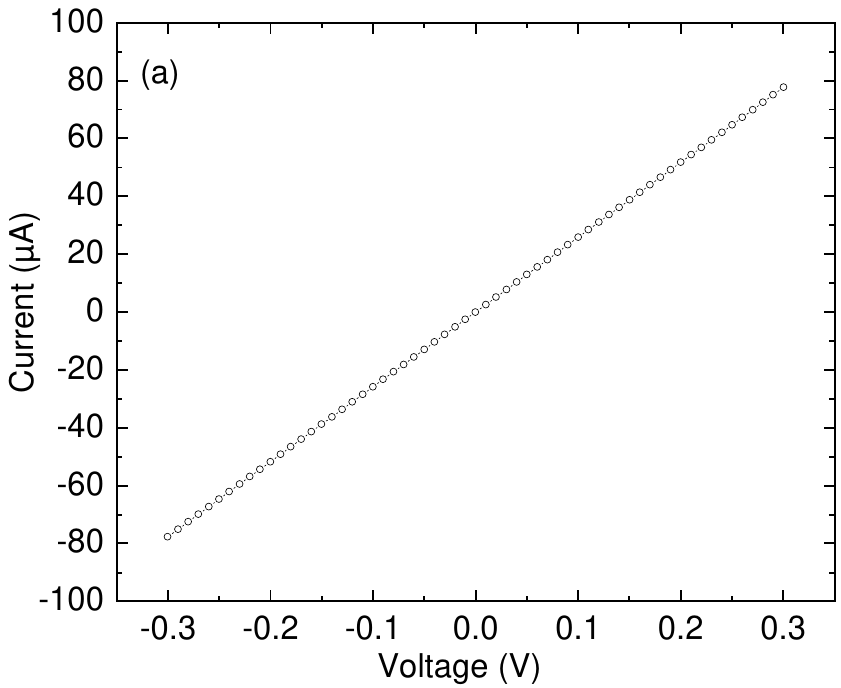}\hspace{0.5cm}\includegraphics[width=6.5cm]{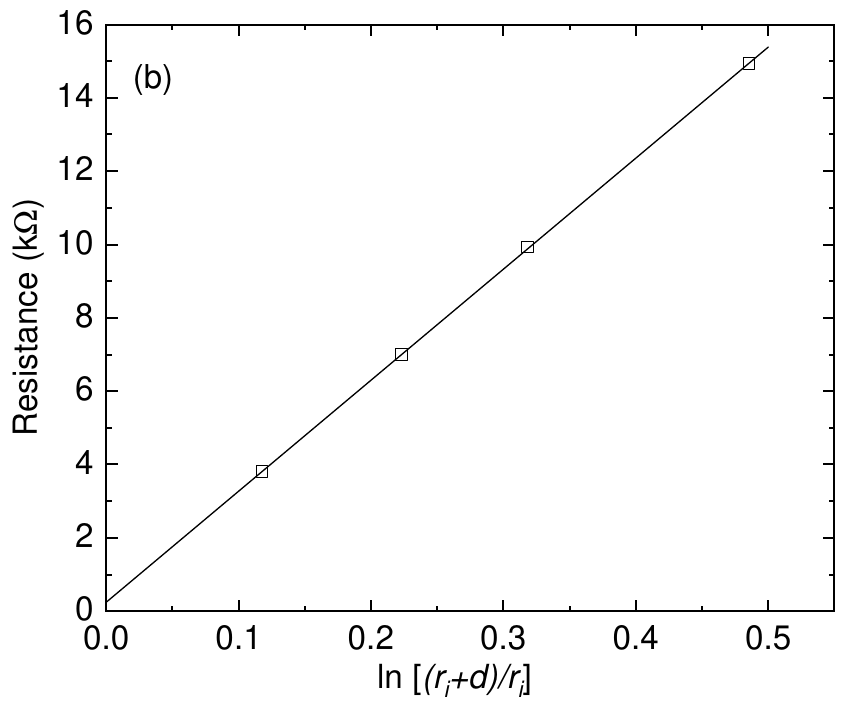}\caption{Characterization of Ti/Au contacts on $p$-SnO in sample A015 using
cTLM patterns. (a) Current-voltage characteristics of the pattern
with lowest spacing (5\,$\mu$m) confirming an ohmic contact to the
SnO. The extracted total resistance of 3.8\,k$\Omega$ and those
from other gap spacings are used in the analysis in (b), which reveals
a sheet resistance (derived from the slope) of 190\,k$\Omega$ and
specific contact resistance of $\approx0.05$\,m$\Omega\thinspace$cm$^{2}$.\label{fig:SIVcTLMA015}}
\end{figure*}
\begin{figure*}
\includegraphics[width=15cm]{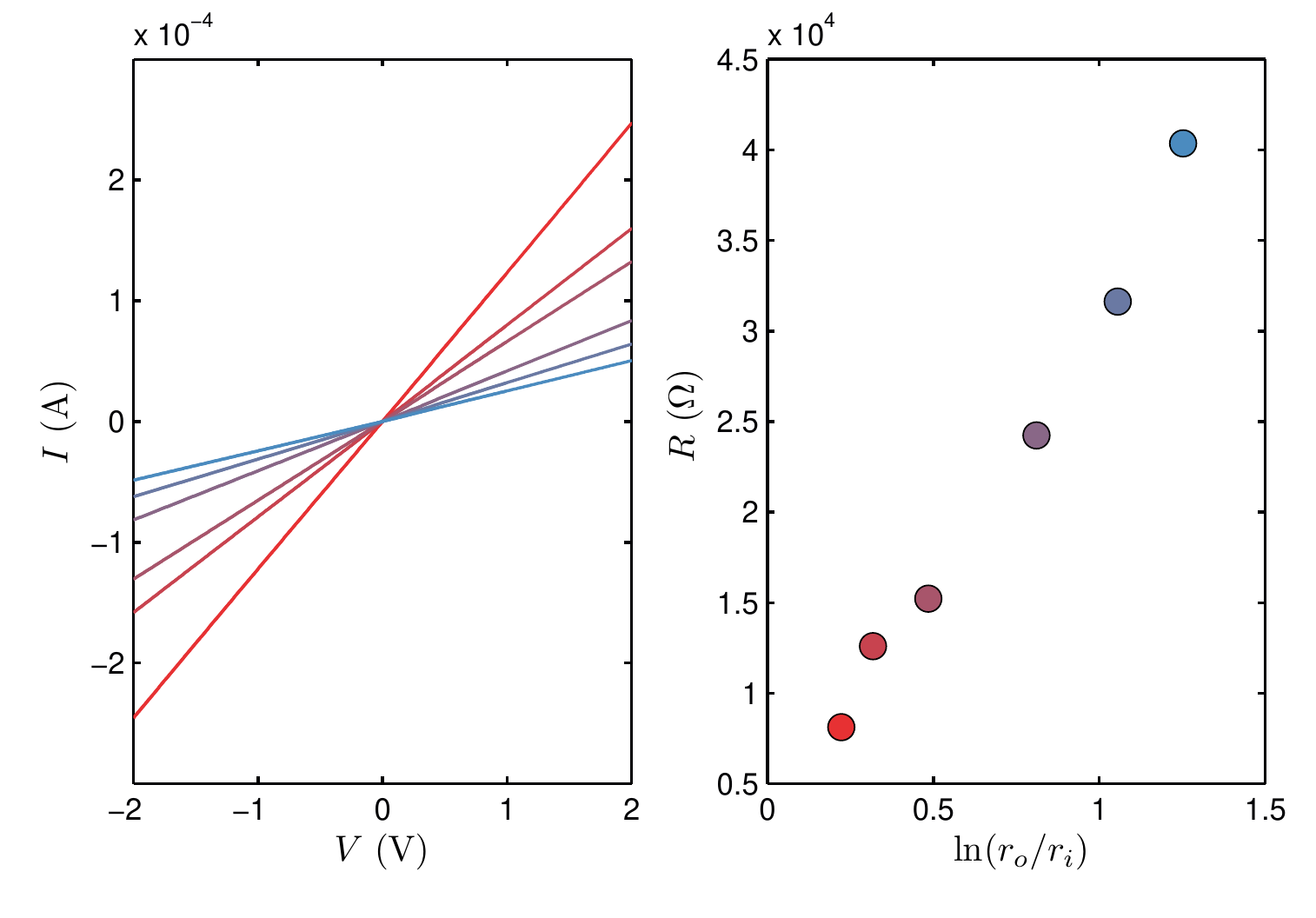}\caption{Characterization of Ti/Au contacts on $p$-SnO in sample G016 using
cTLM patterns. (left) Current-voltage characteristics of the patterns
confirming an ohmic contact to the SnO and negligible current leakage
into the underlying Ga$_{2}$O$_{3}$.\protect \\
(right) The extracted total resistance of from the different gap spacings
are used in the analysis to obtain a sheet resistance (derived from
the slope) of 173\,k$\Omega$ and specific contact resistance of
$\approx3.4$\,m$\Omega\thinspace$cm$^{2}$.\label{fig:SIVcTLMG016}}
\end{figure*}

To assess the Ti/Au contact to $p$-type SnO, used as top contact
in the diodes, we processed circular TLM (cTLM)\citep{Reeves1980}
patterns with inner disc radius of $r_{i}=80\thinspace\mu$m and gap
spacings ranging from $d=$5 to 100\,\textmu m on samples A015 and
G016 using the identical processing scheme to the top contacts of
the diodes (photolithography, electron-beam evaporation, lift-off,
no contact annealing). An optical micrograph of the cTLM structure
(on a SnO$_{2}$ layer) can be found in Ref.\,\citenum{Bierwagen2009a}.
Current-voltage curves of the cTLM structures were linear as shown
exemplarily for the pattern with the lowest gap spacing, i.e., the
one with strongest contribution from the contact, of A015 in Fig.\,\ref{fig:SIVcTLMA015}(a)
and shown for all measured gap spacings of G016 in Fig.\,\ref{fig:SIVcTLMG016}(right).
This result confirms an ohmic contact between Ti and the $p$-type
SnO of both samples. More importantly, the good linearity on G016
indicates negligible leakage of the current across the (rectifying)
$pn$-junction. Analysis of the measured resistances as function of
gap spacing, shown in Fig.\,\ref{fig:SIVcTLMA015}(b) and Fig.\,\ref{fig:SIVcTLMG016}(left),
reveals a sheet resistance of of 190\,k$\Omega$ for A015 and 173\,k$\Omega$
for G016, both in excellent agreement with the values extracted from
the van-der-Pauw measurements. The extracted specific contact resistance
is $\rho_{c}=0.05$\,m$\Omega$cm$^{2}$ and $\rho_{c}=3.4$\,m$\Omega$cm$^{2}$
for A015 and G016, respectively.

\section*{Reverse leakage on G015 and G016 in absence of mesa isolation}

Without mesa etching current can laterally spread from the top contact
pad in the SnO layer (with sheet resistance for lateral transport
on the order of $100$\,k$\Omega$) if it helps minimizing the total
resistance to the back-contact on the Ga$_{2}$O$_{3}$ substrate.
In forward bias, vertical current flow from the $p$-type SnO into
$n$-type Ga$_{2}$O$_{3}$ can take place and the measured series
resistance on the order or 10--100\,$\Omega$ is well below the
sheet resistance of the SnO layer, making current spreading not favorable.
This scenario changes drastically in reverse bias where current flow
across the $pn$-junction is impeded and current spreading in the
SnO helps maximizing the active junction area (or allowing the current
to flow through isolated leakage spots) and thus total current flow
across the $pn$-junction. From a linear fit of the reverse current
in the diodes prior to mesa etching a parallel leakage resistance
on the order of 80\,k$\Omega$ was extracted, which is on the order
of the SnO sheet resistance, strongly suggesting current spreading
to limit the reverse leakage current.

Two successive mesa etchings of G016 to total etch depths of 140 and
180\,nm were performed with significant reduction of the reverse
leakage only at 180\,nm. Consequently, current spreading still took
place after the 140\,nm-etch and thus, the SnO layer thickness on
G016 is between 140 and 180\,nm, in good agreement with the SnO film
thickness of 170\,nm on the reference sample A016.

\section*{Analysis of Laterally inhomogeneous barriers}

\begin{figure*}
\includegraphics[width=8.5cm]{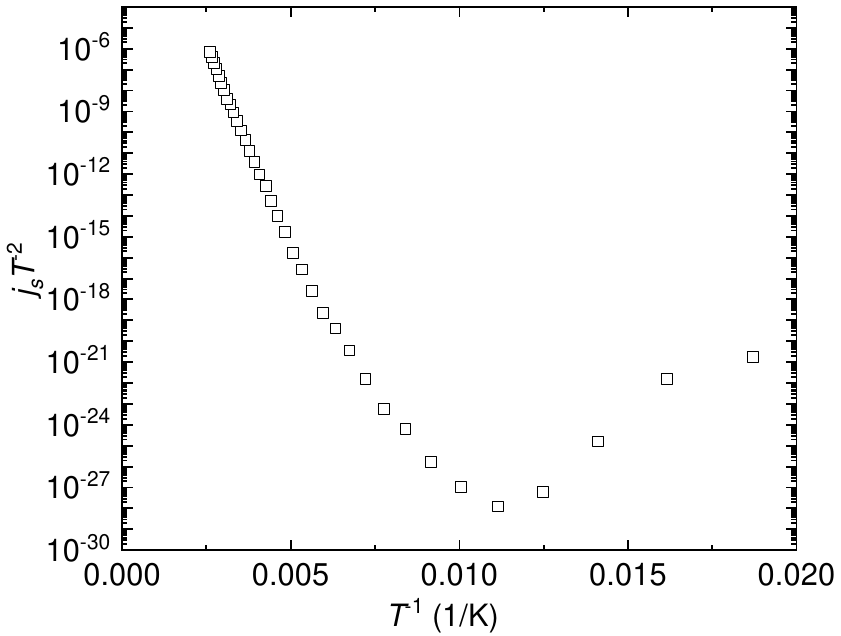}\includegraphics[width=8.5cm]{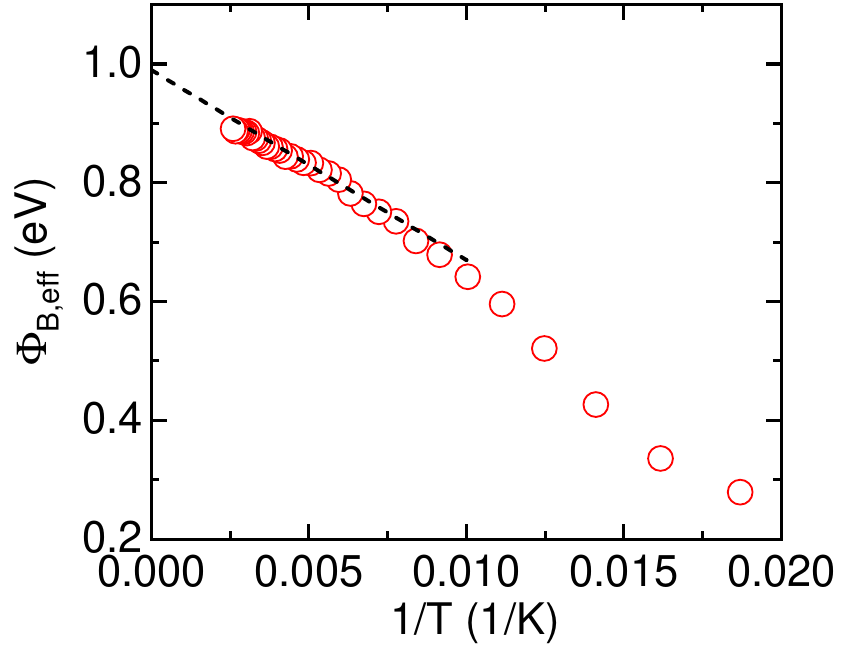}\caption{left: Representative temperature dependence of the saturation current:
$\log(I_{\mathrm{S}}/T\text{\texttwosuperior})$ vs. $T^{-1}$ of
a diode on sample G016.\protect \\
right: Representative linear fit of the effective barrier height $\Phi_{\mathrm{B}}^{\mathrm{eff}}$
to obtain the mean barrier height $\overline{\Phi}_{\mathrm{B,0}}$
and the standard deviation of its distribution $\sigma_{0}$.\label{fig:IsT}}
\end{figure*}
The quadratic behavior $\log(I_{\mathrm{S}}/T\text{\texttwosuperior})$
vs. $T^{-1}$ shown in Fig.\,\ref{fig:IsT}(left) indicates thermionic
emission over a Gaussian distributed lateral inhomogeneous barrier
being the dominant transport mechanism.\citep{Werner1991} This type
of barrier can be characterized by the mean barrier height $\overline{\Phi}_{\mathrm{B,0}}=1.02\pm0.04\,\mathrm{eV}$
and the standard deviation of its distribution $\sigma_{\mathrm{0}}=0.080\pm0.008\,\mathrm{eV}$,
which were extracted from the temperature dependence according to
$\Phi_{\mathrm{B}}^{\mathrm{eff}}=\overline{\Phi}_{\mathrm{B,0}}-\frac{\sigma_{\mathrm{0}}^{2}}{2k_{\mathrm{B}}T}$.\citep{Werner1991}
Note that the values given here and in the following are the mean
values of $IVT$-measurements on four different contacts. Further,
from a linear fit of $\eta^{-1}-1$ vs. $T^{-1}$ (not shown) we determined
the voltage coefficients $\rho_{2}$ and $\rho_{3}$ of $\overline{\Phi}{}_{\mathrm{B,0}}$
and $\sigma_{0}$ to be $-0.15\pm0.04$ and $-0.009\pm0.003\,\mathrm{eV}$,
respectively. Alternatively, the homogeneous barrier height can be
extracted from the empiric model of Schmitsdorf and M�nch \citep{Schmitsdorf1997,Schmitsdorf1999}
by linear extrapolation of $\Phi_{\mathrm{B,eff}}$ vs. $\eta$ to
$\eta=1.01$. The resulting homogeneous barrier height of $0.96\pm0.04\,\mathrm{eV}$
is again in good agreement with the built-in voltage $V_{bi}=1.07\pm0.03$\,V
as well as the mean barrier height.

\bibliographystyle{plainnat}
\bibliography{bibs/Diodes}